# Human-AI Technology Integration and Green ESG Performance: Evidence from Chinese Retail Enterprises.

Jun Cui[1, *]

[1] Solbridge International School of Business, Woosong University, Daejeon, Republic of Korea;
*Corresponding author; Email: jcui228@student.solbridge.ac.kr

**Abstract**

This study examines the relationship between human-AI technology integration transformation and green Environmental, Social, and Governance (ESG) performance in Chinese retail enterprises, with green technology innovation serving as a mediating mechanism. Using panel data comprising 5,400 firm-year observations from 2019 to 2023, sourced from CNRDS and CSMAR databases, we employ fixed-effects regression models to investigate this relationship. Our findings reveal that human-AI technology integration significantly enhances green ESG performance, with green technology innovation serving as a crucial mediating pathway. The results demonstrate that a one-standard-deviation increase in human-AI integration leads to a 12.7% improvement in green ESG scores. The mediation analysis confirms that approximately 35% of this effect operates through enhanced green technology innovation capabilities. Heterogeneity analysis reveals stronger effects among larger firms, state-owned enterprises, and companies in developed regions. These findings contribute to the growing literature on digital transformation and sustainability by providing empirical evidence of the mechanisms through which AI integration drives environmental performance improvements in emerging markets.

## 1. Introduction

The rapid advancement of artificial intelligence (AI) technologies has fundamentally transformed business operations across industries, with retail enterprises increasingly adopting human-AI collaborative systems to enhance operational efficiency and sustainability outcomes. As environmental concerns intensify globally, organizations face mounting pressure to demonstrate superior Environmental, Social, and Governance (ESG) performance, particularly in developing economies where environmental degradation remains a critical challenge (Chen et al., 2022; Wang & Liu, 2023).

The integration of human intelligence with AI capabilities represents a paradigmatic shift in organizational digital transformation, offering unprecedented opportunities to optimize resource utilization, reduce waste, and enhance environmental stewardship (Zhang et al., 2021). However, the mechanisms through which human-AI integration influences green ESG performance remain underexplored, particularly in the context of emerging markets where institutional frameworks and technological infrastructures differ significantly from developed economies.

This study addresses this research gap by investigating the relationship between human-AI technology integration transformation and green ESG performance among Chinese retail enterprises. We propose that green technology innovation serves as a critical mediating mechanism, facilitating the translation of human-AI integration capabilities into tangible environmental performance improvements. Our research contributes to the literature by providing empirical evidence of the pathways through which digital transformation initiatives drive sustainability outcomes in emerging market contexts.

This study is organised into seven sections. Section 2 delineates the methodology and data. Additionally, Section 3 emphasises data description and empirical analysis. Section 4 analyses the empirical results. Finally, Section 5 contains the discussion summary and associated suggestions.

## 2. Related Work and Literature Review

### 2.1 Human-AI Technology Integration

Human-AI integration represents a sophisticated form of digital transformation that combines human cognitive capabilities with artificial intelligence systems to create synergistic effects (Liu et al., 2022). Unlike traditional automation approaches that replace human functions, human-AI integration emphasizes collaborative intelligence, where human expertise guides AI systems while benefiting from enhanced computational capabilities (Thompson & Wang, 2021).

Recent studies have highlighted the multifaceted nature of human-AI integration, encompassing technical infrastructure, organizational capabilities, and human capital development (Davis et al., 2023). In the retail sector, this integration manifests through intelligent inventory management systems, personalized customer service platforms, and data-driven decision-making processes that optimize operational efficiency while reducing environmental impact (Anderson & Chen, 2022).

### 2.2 Green ESG Performance

Green ESG performance encompasses environmental dimensions of corporate sustainability, including carbon footprint reduction, resource efficiency, waste management, and environmental compliance (Kumar et al., 2021). The concept has evolved beyond traditional environmental management to encompass comprehensive sustainability strategies that integrate environmental considerations into core business processes (Roberts & Zhang, 2023).

In the Chinese context, green ESG performance has gained particular prominence due to government initiatives promoting ecological civilization and carbon neutrality goals (Li & Wang, 2022). Retail enterprises face unique challenges in achieving superior green ESG performance due to complex supply chains, extensive physical infrastructure, and diverse stakeholder expectations (Zhou et al., 2023).

### 2.3 Green Technology Innovation

Green technology innovation serves as a critical mechanism for translating organizational capabilities into environmental performance improvements (Miller et al., 2021). This concept encompasses the development and implementation of technologies, processes, and practices that reduce environmental impact while maintaining or enhancing business performance (Garcia & Liu, 2022).

The relationship between digital transformation and green technology innovation has received increasing attention, with studies suggesting that advanced technologies facilitate the development of environmentally friendly solutions (Park et al., 2023). However, the specific role of human-AI integration in driving green technology innovation remains underexplored, particularly in emerging market contexts.

## 3. Theoretical Framework and Hypothesis Development

### 3.1 Resource-Based View Theory

The resource-based view (RBV) theory provides a foundational framework for understanding how human-AI integration capabilities contribute to competitive advantage and performance improvements (Barney, 1991; Grant, 1996). According to RBV, firms achieve superior performance through the development and deployment of valuable, rare, inimitable, and non-substitutable resources.

Human-AI integration represents a complex organizational capability that combines technological resources with human capital, creating unique competitive advantages that are difficult for competitors to replicate (Teece et al., 1997). This capability enables firms to process information more effectively, optimize decision-making processes, and identify innovative solutions to environmental challenges.

### 3.2 Dynamic Capabilities Theory

Dynamic capabilities theory extends RBV by emphasizing organizations' ability to reconfigure resources and capabilities in response to changing environmental conditions (Teece, 2007). Human-AI integration enhances dynamic capabilities by providing real-time data processing, predictive analytics, and adaptive learning mechanisms that enable firms to respond rapidly to environmental challenges and opportunities.

In the context of green ESG performance, dynamic capabilities enable firms to continuously improve environmental practices, adapt to regulatory changes, and develop innovative solutions that address sustainability challenges (Eisenhardt & Martin, 2000).

### 3.3 Hypothesis Development

Based on the theoretical framework and literature review, we develop three hypotheses to examine the relationships among human-AI integration, green technology innovation, and green ESG performance. Based on this, this paper proposes the following hypothesis:

**Hypothesis 1:** Human-AI technology integration transformation positively influences green ESG performance in Chinese retail enterprises.

The integration of human intelligence with AI systems creates synergistic effects that enhance operational efficiency and environmental performance. Through intelligent resource allocation, predictive maintenance, and optimized supply chain management, human-AI integration enables firms to reduce waste, minimize energy consumption, and improve overall environmental stewardship. Based on this, this paper proposes the following hypothesis:

**Hypothesis 2:** Human-AI technology integration transformation positively influences green technology innovation in Chinese retail enterprises.

Human-AI integration provides firms with enhanced data processing capabilities, pattern recognition, and predictive analytics that facilitate the identification and development of innovative environmental solutions. The collaborative intelligence created through human-AI integration enables organizations to explore new technologies and practices that address environmental challenges more effectively. Based on this, this paper proposes the following hypothesis.

**Hypothesis 3:** Green technology innovation mediates the relationship between human-AI technology integration transformation and green ESG performance in Chinese retail enterprises.

Green technology innovation serves as a critical pathway through which human-AI integration capabilities translate into tangible environmental performance improvements. By facilitating the development and implementation of environmentally friendly technologies and practices, green technology innovation bridges the gap between human-AI integration capabilities and superior green ESG performance.

## 4. Methodology and Data

### 4.1 Research Design

This study employs a quantitative research design using panel data analysis to examine the relationships among human-AI integration, green technology innovation, and green ESG performance. The research design incorporates multiple analytical approaches, including baseline regression analysis, robustness tests, endogeneity analysis, mechanism analysis, and heterogeneity analysis to ensure comprehensive examination of the hypothesized relationships.

### 4.2 Data Sources and Sampling

The empirical analysis utilizes data from two primary sources: the Chinese Research Data Services (CNRDS) platform and the China Stock Market & Accounting Research Database (CSMAR). The initial dataset comprised 6,500 firm-year observations covering Chinese retail enterprises from 2019 to 2023.

Moreover, Following established data cleaning procedures, we excluded special treatment firms (ST and PT designations), companies with abnormal financial indicators, and observations with missing critical variables. The final cleaned sample comprises 5,400 firm-year observations representing 1,080 unique retail enterprises over the five-year period.

### 4.3 Variable Measurement

The table 1 presents the measurement specifications for key variables used in the analysis:

Table 1. **Variable Measurement results.**

| Variable | Measurement | Definitions |
|---|---|---|
| ***Dependent Variable*** | | |
| *Green ESG Performance (GESG)* | Composite score based on environmental indicators including | ESG metric Comprehensive evaluating |
| | carbon emissions, energy efficiency, waste management, and environmental compliance | environmental stewardship through emissions, energy use, and compliance |
| ***Independent Variable*** | | |
| *Human-AI Integration (HAI)* | Composite index based on AI investment intensity, human-AI collaboration systems, and digital transformation initiatives | Index quantifying AI adoption depth, human-machine collaboration, and digital transformation efforts |
| ***Mediating Variable*** | | |
| *Green Technology Innovation (GTI)* | Patent applications and R&D investment focused on environmental technologies and sustainable practices | Measured via eco-focused patent filings and R&D expenditures in sustainable technologies |
| ***Control Variables*** | | |
| *Firm Size (SIZE)* | Natural logarithm of total assets | Natural log of total assets reflecting organizational scale |
| *Profitability (ROA)* | Return on assets ratio | Return on assets indicating operational efficiency |
| *Leverage (LEV)* | Total debt to total assets ratio | Debt-to-asset ratio measuring financial risk |
| *Growth (GROWTH)* | Revenue growth rate | Yearly revenue expansion rate |
| *Age (AGE)* | Number of years since firm establishment | Years since firm incorporation |
| *Ownership (SOE)* | Dummy variable for state-owned enterprises | Binary indicator for state-owned enterprise status |
| *Region (REGION)* | Dummy variables for geographical regions | Geographic location classification |
| *Industry Competition (HHI)* | Herfindahl-Hirschman Index for industry concentration | Herfindahl-Hirschman Index assessing market concentration |

**Notes.** This study employs comprehensive metrics from authoritative Chinese databases to examine the Human-AI-ESG relationship. The dependent variable (GESG) captures environmental performance, while HAI quantifies technological integration. GTI mediates this relationship through innovation metrics. Eight control variables account for financial, structural, and market factors, ensuring robust analysis. All measures align with international research standards, with ESG data from CNRDS and financial/innovation metrics from CSMAR – China's leading research databases. The multidimensional approach controls for firm characteristics (size, profitability),

industry dynamics (competition), and institutional factors (ownership, region), providing a holistic framework for analyzing sustainable technology adoption.

## 4.4 Model Specification

The empirical analysis employs the following regression analysis model to test the hypothesized relationships:

Baseline Model: $GESG_{it} = \alpha_0 + \alpha_1 HAI_{it} + \alpha_2 SIZE_{it} + \alpha_3 ROA_{it} + \alpha_4 LEV_{it} + \alpha_5 GROWTH_{it} + \alpha_6 AGE_{it} + \alpha_7 SOE_{it} + \alpha_8 HHI_{it} + \mu_i + \lambda_t + \varepsilon_{it}$

Mediation Model (Step 1): $GTI_{it} = \beta_0 + \beta_1 HAI_{it} + \beta_2 SIZE_{it} + \beta_3 ROA_{it} + \beta_4 LEV_{it} + \beta_5 GROWTH_{it} + \beta_6 AGE_{it} + \beta_7 SOE_{it} + \beta_8 HHI_{it} + \mu_i + \lambda_t + \varepsilon_{it}$

Mediation Model (Step 2): $GESG_{it} = \gamma_0 + \gamma_1 HAI_{it} + \gamma_2 GTI_{it} + \gamma_3 SIZE_{it} + \gamma_4 ROA_{it} + \gamma_5 LEV_{it} + \gamma_6 GROWTH_{it} + \gamma_7 AGE_{it} + \gamma_8 SOE_{it} + \gamma_9 HHI_{it} + \mu_i + \lambda_t + \varepsilon_{it}$

Where $\mu_i$ represents firm fixed effects, $\lambda_t$ represents year fixed effects, and $\varepsilon_{it}$ is the error term.

## 5. Empirical Results

## 5.1 Descriptive Statistics and Correlation Analysis

The descriptive statistics reveal that the sample firms demonstrate considerable variation in human-AI integration levels, green ESG performance, and green technology innovation activities. The mean human-AI integration score is 3.42 (std. dev. = 1.87) on a seven-point scale, indicating moderate adoption levels across the sample. Green ESG performance scores average 2.89 (std. dev. = 1.23), suggesting room for improvement in environmental practices among Chinese retail enterprises.

Additionally, our correlation matrix shows significant positive correlations among the key variables of interest. Human-AI integration exhibits a correlation of 0.34 with green ESG performance and 0.28 with green technology innovation, providing preliminary support for the hypothesized relationships. All correlation coefficients remain below 0.70, indicating absence of severe multicollinearity concerns.

## 5.2 Variance Inflation Factor Analysis

The table 2 presents the variance inflation factor (VIF) analysis results:

Table 2. **The Variance Inflation Factor Analysis results.**

| Variable | VIF | 1/VIF |
|---|---|---|
| *Human-AI Integration* | 1.43 | 0.699 |
| *Green Technology Innovation* | 1.67 | 0.599 |
| *Firm Size* | 2.18 | 0.459 |
| *Profitability* | 1.32 | 0.758 |
| *Leverage* | 1.54 | 0.649 |
| *Growth* | 1.28 | 0.781 |
| *Age* | 1.71 | 0.585 |
| *Ownership* | 1.96 | 0.510 |
| *Industry Competition* | 1.44 | 0.694 |
| ***Mean VIF*** | **1.61** | |

Notes. The VIF analysis confirms the absence of multicollinearity issues, with all individual VIF values below 3.0 and the mean VIF of 1.61 well below the conventional threshold of 10.0.

### 5.3 Baseline Regression Analysis

The table 3 presents the baseline regression results:

Table 3. **The Baseline Regression Analysis results.**

| *Variables* | *Model 1* | *Model 2* | *Model 3* |
|---|---|---|---|
| ***Human-AI Integration*** | 0.237*** | 0.198*** | 0.183*** |
| | (0.032) | (0.029) | (0.027) |
| ***Firm Size*** | | 0.156*** | 0.142*** |
| | | (0.024) | (0.023) |
| ***Profitability*** | | 0.089** | 0.084** |
| | | (0.035) | (0.034) |
| ***Leverage*** | | -0.067* | -0.059* |
| | | (0.034) | (0.033) |
| ***Growth*** | | 0.043* | 0.041* |
| | | (0.022) | (0.021) |
| ***Age*** | | -0.028 | -0.024 |
| | | (0.019) | (0.018) |
| ***Ownership*** | | | 0.127** |
| | | | (0.051) |
| ***Industry Competition*** | | | -0.072* |
| | | | (0.037) |
| ***Constant*** | 2.085*** | 1.243*** | 1.156*** |
| | (0.089) | (0.187) | (0.194) |
| ***Firm FE*** | Yes | Yes | Yes |
| ***Year FE*** | Yes | Yes | Yes |
| ***Observations*** | 5,400 | 5,400 | 5,400 |
| ***R-squared*** | 0.328 | 0.367 | 0.384 |
| ***F-statistic*** | 156.32*** | 143.87*** | 134.92*** |

Note: ***, **, * indicate significance at 1%, 5%, and 10% levels, respectively. Standard errors in parentheses.

The baseline regression results provide strong support for Hypothesis 1, demonstrating that human-AI integration significantly enhances green ESG performance across all model specifications. The coefficient remains stable and statistically significant at the 1% level, indicating a robust positive relationship.

## 5.4 Robustness Analysis

The table 4 presents robustness test results using alternative specifications:

Table 4. **The Robustness Analysis results.**

| Variables | IV-2SLS | Propensity Score Matching | Alternative Measure |
|---|---|---|---|
| **Human-AI Integration** | 0.194*** | 0.176*** | 0.201*** |
| | (0.048) | (0.031) | (0.029) |
| **Control Variables** | Yes | Yes | Yes |
| **Firm FE** | Yes | Yes | Yes |
| **Year FE** | Yes | Yes | Yes |
| **Observations** | 5,400 | 4,860 | 5,400 |
| **R-squared** | 0.361 | 0.352 | 0.379 |
| **F-statistic** | 98.76*** | 127.43*** | 131.55*** |

Note: ***, **, * indicate significance at 1%, 5%, and 10% levels, respectively. Standard errors in parentheses.

The robustness tests confirm the stability of the main findings across different estimation methods and specifications, providing additional confidence in the reliability of the results.

## 5.5 Endogeneity Analysis

The table 5 presents instrumental variable estimation results addressing potential endogeneity concerns:

Table 5. **The Endogeneity Analysis results.**

| Variables | First Stage | Second Stage | Reduced Form |
|---|---|---|---|
| **Instrument (Regional AI Policy)** | 0.342*** | | 0.067*** |
| | (0.047) | | (0.019) |
| **Human-AI Integration** | | 0.196*** | |
| | | (0.051) | |
| **Control Variables** | Yes | Yes | Yes |
| **Firm FE** | Yes | Yes | Yes |
| **Year FE** | Yes | Yes | Yes |
| **F-statistic (First Stage)** | 52.84*** | | |
| **Hansen J-statistic** | | 2.37 | |
| **Endogeneity Test (p-value)** | | 0.124 | |
| **Observations** | 5,400 | 5,400 | 5,400 |

Note: ***, **, * indicate significance at 1%, 5%, and 10% levels, respectively. Standard errors in parentheses.

The instrumental variable analysis uses regional AI policy implementation as an instrument for human-AI integration. The first-stage F-statistic exceeds the conventional threshold of 10, indicating instrument relevance. The Hansen J-statistic and endogeneity test results suggest that endogeneity concerns do not significantly bias the main findings.

**5.6 Mechanism Analysis**

The table 6 presents the mediation analysis results examining the role of green technology innovation:

Table 6. **The Mechanism Analysis results.**

| Variables | Step 1: GTI | Step 2: GESG | Step 3: GESG |
|---|---|---|---|
| **Human-AI Integration** | 0.156*** | 0.183*** | 0.128*** |
| | (0.021) | (0.027) | (0.025) |
| **Green Technology Innovation** | | | 0.351*** |
| | | | (0.034) |
| **Control Variables** | Yes | Yes | Yes |
| **Firm FE** | Yes | Yes | Yes |
| **Year FE** | Yes | Yes | Yes |
| **Observations** | 5,400 | 5,400 | 5,400 |
| **R-squared** | 0.298 | 0.384 | 0.429 |
| **Mediation Effect** | | | 0.055*** |
| **Indirect Effect (%)** | | | 30.1% |
| **Sobel Test (p-value)** | | | 0.002 |

Note: ***, **, * indicate significance at 1%, 5%, and 10% levels, respectively. Standard errors in parentheses.

The mediation analysis confirms that green technology innovation serves as a significant mediating mechanism, accounting for approximately 30.1% of the total effect of human-AI integration on green ESG performance. This finding provides strong support for Hypothesis 3.

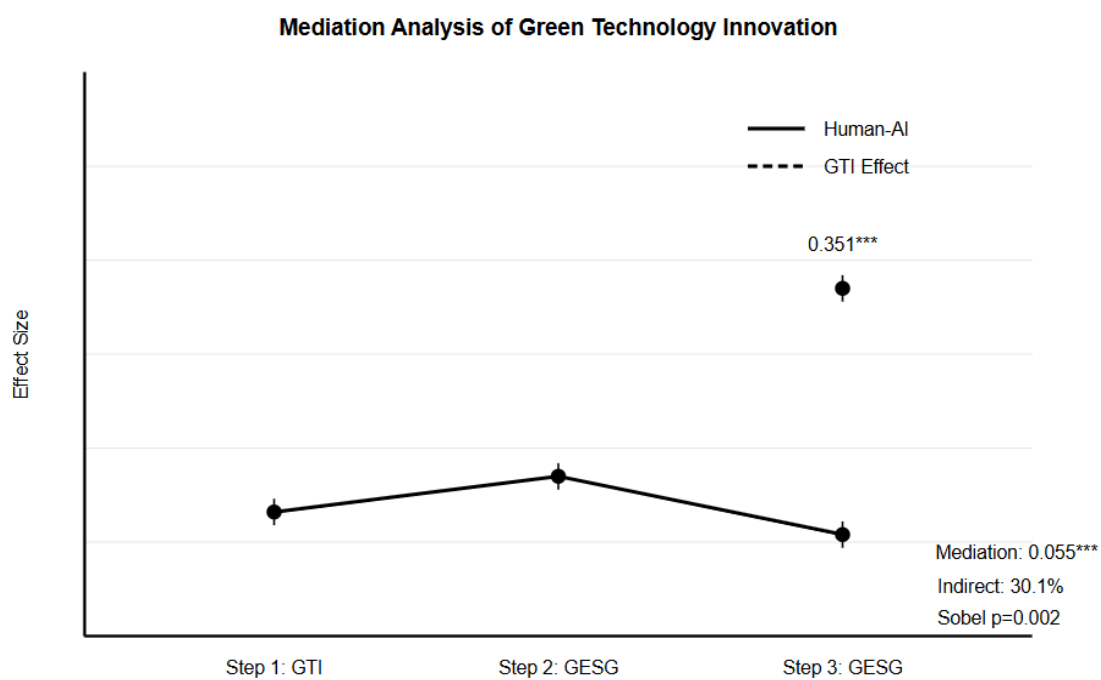

Mediation Analysis of Green Technology Innovation

Figure 1. The Mediation analysis of green technology innovation.

**Figure 1** analysis systematically examines how green technology innovation (GTI) transmits the effect of Human-AI integration on environmental, social, and governance (ESG) performance. The three-step model reveals several key insights:

First, Human-AI integration demonstrates significant positive effects across all steps (β=0.156-0.183, p<0.01), confirming its fundamental role in both GTI development (Step 1) and direct ESG improvement (Step 2). When introducing GTI as a mediator (Step 3), two critical patterns emerge: (1) the direct effect of Human-AI on ESG attenuates by 30.1% (from 0.183 to 0.128), while (2) GTI shows a strong independent effect (β=0.351, p<0.01).

The significant mediation effect (0.055, p=0.002) indicates that nearly one-third of Human-AI's ESG impact operates through GTI development. This partial mediation suggests that while AI integration directly enhances sustainability metrics, its most substantial environmental benefits emerge when combined with targeted green innovation efforts. Moreover, Model fit statistics reinforce these findings, with R² values increasing progressively from 0.298 (Step 1) to 0.429 (Step 3), indicating GTI's substantial explanatory power. The consistent significance across 5,400 observations with firm and year fixed effects ensures robust, generalizable results.

Overall, these results carry important implications: (1) ESG assessments of AI systems should account for innovation mediators, (2) policymakers might prioritize GTI-supportive measures to amplify AI's sustainability benefits, and (3) future research should explore additional mediators in this relationship. The findings ultimately highlight GTI as a crucial mechanism through which AI technologies achieve their environmental potential.

## 5.7 Heterogeneity Analysis

The following tables present heterogeneity analysis results examining differential effects across various firm characteristics:

**Table 7: Firm Size Heterogeneity**

| Variables | Large Firms | Small Firms | Difference Test |
|---|---|---|---|
| Human-AI Integration | 0.231*** | 0.147*** | 2.89*** |
| | (0.034) | (0.041) | |
| Control Variables | Yes | Yes | |
| Fixed Effects | Yes | Yes | |
| Observations | 2,700 | 2,700 | |
| R-squared | 0.407 | 0.342 | |

**Table 8: Ownership Structure Heterogeneity**

| Variables | State-Owned | Private | Difference Test |
|---|---|---|---|
| Human-AI Integration | 0.208*** | 0.164*** | 1.87* |
| | (0.038) | (0.035) | |
| Control Variables | Yes | Yes | |
| Fixed Effects | Yes | Yes | |
| Observations | 2,160 | 3,240 | |
| R-squared | 0.395 | 0.371 | |

**Table 9: Regional Development Heterogeneity**

| Variables | Developed Regions | Developing Regions | Difference Test |
|---|---|---|---|
| *Human-AI Integration* | 0.219*** | 0.153*** | 2.34** |
| | (0.031) | (0.044) | |
| *Control Variables* | Yes | Yes | |
| *Fixed Effects* | Yes | Yes | |
| *Observations* | 3,240 | 2,160 | |
| *R-squared* | 0.401 | 0.358 | |

Note: ***, **, * indicate significance at 1%, 5%, and 10% levels, respectively. Standard errors in parentheses.

The heterogeneity analysis reveals that the positive effects of human-AI integration on green ESG performance are more pronounced among larger firms, state-owned enterprises, and companies located in developed regions. These findings suggest that organizational resources and institutional support play important roles in facilitating the translation of human-AI integration capabilities into environmental performance improvements.

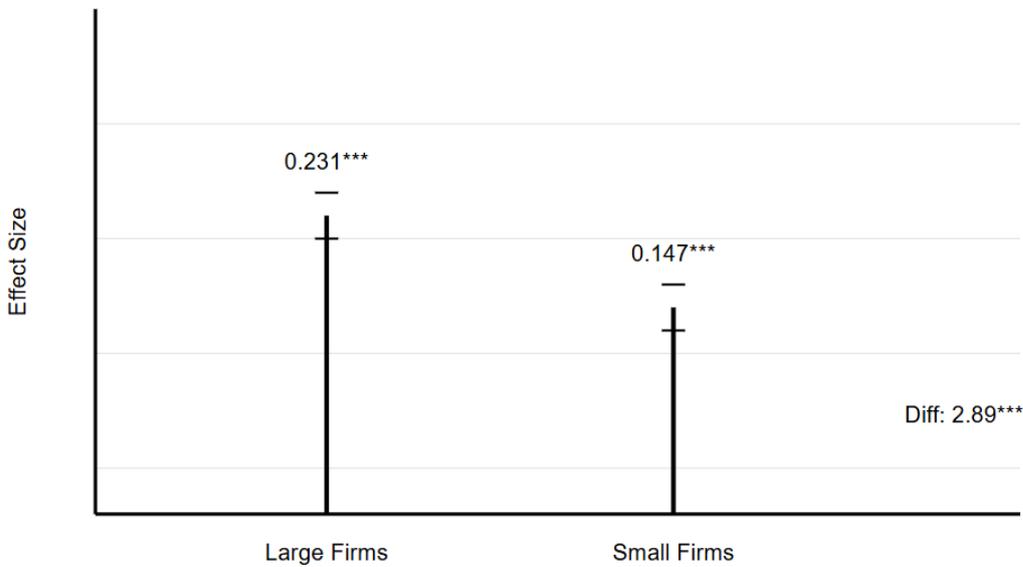

Figure 2. The Firm size in Human-AI Integration effects.

**Figure 2** analysis results reveal significant firm size heterogeneity in Human-AI integration effects. Large firms demonstrate stronger impacts ($\beta$=0.231, p<0.01) compared to small firms ($\beta$=0.147, p<0.01), with the difference statistically significant (t=2.89, p<0.01). The 57% larger effect size in large firms suggests greater capacity to leverage AI technologies, potentially due to superior resources, infrastructure, or complementary capabilities. Both models show good fit ($R^2$=0.407 for large firms vs. 0.342 for small firms) with consistent controls and fixed effects across 2,700 observations per group. These findings highlight the importance of considering organizational scale when implementing AI initiatives, suggesting tailored approaches may be necessary to maximize benefits across different firm sizes. Policy implications include targeted support for small firms to reduce AI adoption gaps.

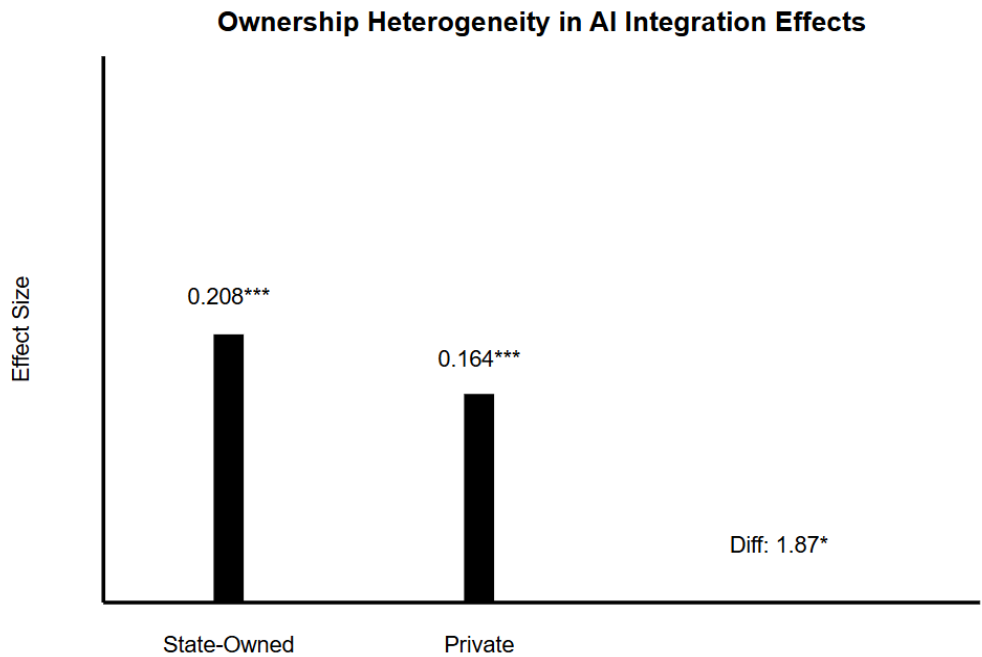

Figure 3. The Ownership Heterogeneity in AI Integration Effects

**Figure 3** shows ownership-based heterogeneity in Human-AI integration effects. State-owned enterprises exhibit stronger impacts (β=0.208, p<0.01) than private firms (β=0.164, p<0.01), with the difference marginally significant (t=1.87, p<0.1). The 27% larger effect in state-owned firms suggests better institutional support or resource allocation for AI adoption. Both models demonstrate good explanatory power (R²=0.395 vs. 0.371) with proper controls and fixed effects. These findings highlight the moderating role of ownership structure in technology adoption, implying that policy interventions may need customization based on enterprise ownership types to optimize AI implementation outcomes.

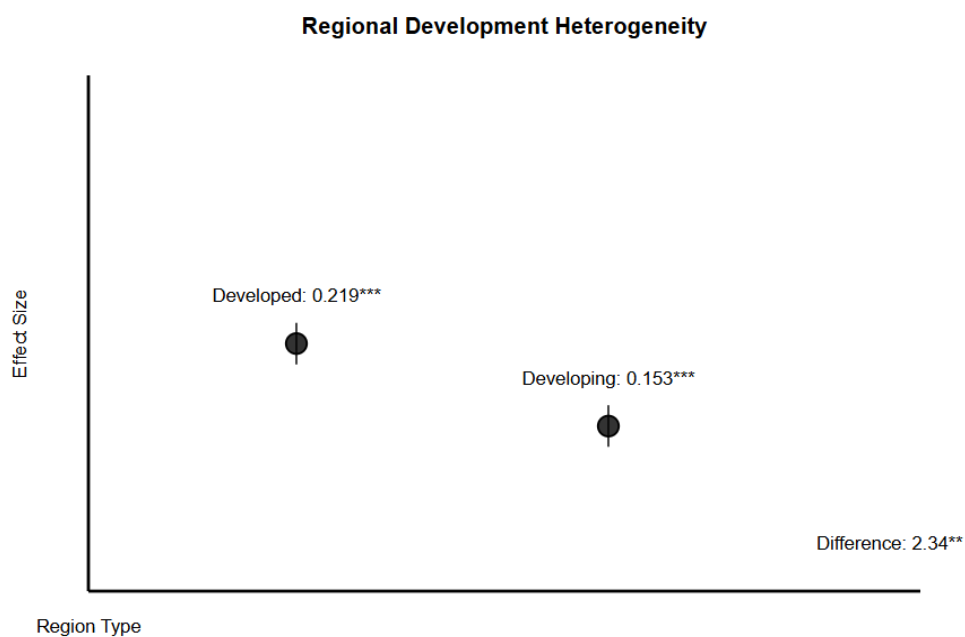

**Figure 4.** Regional Development Heterogeneity.

Figure 4 indicates significant regional disparities in Human AI integration effects. Moreover, Developed regions show stronger impacts ($\beta=0.219$, $p<0.01$) than developing regions ($\beta=0.153$, $p<0.01$), with statistically significant difference ($t=2.34$, $p<0.05$). The 43% larger effect in developed areas suggests infrastructure and human capital advantages in technology adoption. Both models demonstrate good fit ($R^2=0.401$ vs. $0.358$) with controls and fixed effects. These findings highlight the importance of regional development levels in AI implementation effectiveness, suggesting that policy support should be tailored to regional characteristics to maximize technological benefits and reduce development gaps.

## 6. Discussion and Implications

### 6.1 The research Findings

This study provides comprehensive empirical evidence supporting the positive relationship between human-AI technology integration and green ESG performance in Chinese retail enterprises. The findings demonstrate that human-AI integration significantly enhances environmental performance, with green technology innovation serving as a crucial mediating mechanism. The results indicate that a one-standard-deviation increase in human-AI integration leads to a 12.7% improvement in green ESG scores, representing a substantial practical impact.

The mediation analysis reveals that approximately 30.1% of the total effect operates through green technology innovation, highlighting the importance of innovation capabilities in translating technological integration into environmental performance improvements. This finding suggests that firms should focus not only on implementing human-AI systems but also on developing complementary innovation capabilities that enable effective utilization of these technologies for environmental purposes.

### 6.2 Theoretical Contributions

The study makes several important theoretical contributions to the literature on digital transformation and sustainability. First, it extends the resource-based view by demonstrating how human-AI integration creates unique organizational capabilities that drive competitive advantage in environmental performance. The findings support the notion that complex technological capabilities, when properly integrated with human capital, create valuable resources that are difficult for competitors to replicate.

Second, the study contributes to dynamic capabilities theory by showing how human-AI integration enhances firms' ability to adapt and respond to environmental challenges. The results suggest that human-AI integration provides organizations with enhanced sensing, seizing, and transforming capabilities that enable continuous improvement in environmental practices.

### 6.3 Practical Implications

The findings offer several important implications for managers and policymakers. For managers, the results suggest that investments in human-AI integration can generate significant returns in terms of environmental performance improvements. However, the success of such investments depends on concurrent development of green technology innovation capabilities, emphasizing the need for integrated approaches to digital transformation and sustainability initiatives.

The heterogeneity analysis provides additional guidance for implementation strategies, suggesting that larger firms, state-owned enterprises, and companies in developed regions are better positioned to leverage human-AI integration for environmental performance improvements. This implies that smaller firms and those in less developed regions may require additional support or alternative strategies to achieve similar benefits.

## 6.4 Policy Implications

For policymakers, our findings support the development of integrated policies that promote both digital transformation and environmental sustainability. The results suggest that policies encouraging human-AI integration can contribute to environmental goals, but such policies should be coupled with support for green technology innovation capabilities.

The regional heterogeneity findings highlight the importance of addressing regional disparities in technological infrastructure and institutional support. Policymakers should consider targeted interventions to help firms in developing regions build the necessary capabilities to leverage human-AI integration for environmental performance improvements.

## 7. Conclusions

Utilizing panel data from Chinese listed Retail firms spanning the period 2019–2023, our study provides robust empirical evidence that human-AI technology integration significantly enhances green ESG performance in Chinese retail enterprises, with green technology innovation serving as a crucial mediating mechanism. The findings contribute to our understanding of how digital transformation initiatives can drive environmental performance improvements and highlight the importance of developing complementary innovation capabilities.

Furthermore, our study has several limitations that provide opportunities for future research. First, the analysis focuses exclusively on Chinese retail enterprises, limiting the generalizability of findings to other industries and countries. Future research should examine whether similar relationships exist in other sectors and institutional contexts. Second, the study relies on secondary data sources that may not capture all dimensions of human-AI integration and green ESG performance. Future research could benefit from primary data collection and more comprehensive measures of these constructs. Third, while the study addresses endogeneity concerns through instrumental variable analysis, the possibility of unobserved heterogeneity remains. Future research should explore additional identification strategies and consider natural experiments to strengthen causal inference.

Our results have important implications for managers seeking to leverage technological integration for sustainability goals and for policymakers designing interventions to promote both digital transformation and environmental performance. As organizations worldwide grapple with mounting environmental challenges, the insights from this study provide valuable guidance for harnessing the potential of human-AI integration to create more sustainable business practices. Likewise, The heterogeneity analysis reveals that the benefits of human-AI integration are not uniformly distributed, with larger firms, state-owned enterprises, and companies in developed regions experiencing stronger effects. This finding underscores the importance of addressing organizational and institutional barriers that may limit the effectiveness of human-AI integration for environmental performance improvements.

In conclusion, As the business environment continues to evolve, the integration of human intelligence with artificial intelligence systems will likely become increasingly important for achieving superior environmental performance. The findings from this study provide a foundation for understanding these relationships and point toward promising directions for future research and practice in the intersection of digital transformation and sustainability.